\documentclass[11pt, a4paper]{article} 

\usepackage[ansinew]{inputenc}
\usepackage{amsmath, amssymb, graphics, amsthm}
\usepackage{epsfig}
\usepackage{color, xcolor}
\usepackage{fancyhdr} 
\usepackage{manfnt}
\usepackage[T1]{fontenc}

\usepackage[normalem]{ulem}

\newcommand{\w}{\sqrt{1+ q^{cd} \partial_c \phi \partial_d \phi}}

\makeatletter
\@addtoreset{equation}{section}
\makeatother

\newcommand{\be}{\begin{equation}}
\newcommand{\ee}{\end{equation}}
\newcommand{\ba}{\begin{eqnarray}}
\newcommand{\ea}{\end{eqnarray}}

\def\pb#1{\rlap{\lower1.5ex\hbox{$\longleftarrow$}}{#1}}
\def\dpb#1{\rlap{\lower1.5ex\hbox{$\Longleftarrow$}}{#1}}
\def\spb#1{\rlap{\lower1.0ex\hbox{$\leftarrow$}}{#1}}
\def\sdpb#1{\rlap{\lower1.0ex\hbox{$\Leftarrow$}}{#1}}

\title{{\sf On the canonical structure of general relativity with a limiting curvature and its relation to loop quantum gravity }}
\author{
{\sf N. Bodendorfer$^1$\thanks{{\sf 
norbert.bodendorfer@physik.uni-regensburg.de}}, A. Sch\"afer$^2$\thanks{{\sf 
andreas.schaefer@physik.uni-regensburg.de}}, J. Schliemann$^2$\thanks{{\sf john.schliemann@physik.uni-regensburg.de}}}\\
\\
{\sf  ${}^1$~Arnold Sommerfeld Center for Theoretical Physics, LMU Munich,}\\ {\sf Theresienstra{\ss}e 37, 80333 Munich, Germany}\\ \mbox{} \\ {\sf$ {}^2$ Institute for Theoretical Physics, University of Regensburg, } \\ {\sf 93040 Regensburg, Germany  }}
\date{{\small\sf \today}}

\begin{document} 

\maketitle

{\sf

\begin{abstract}

Chamseddine and Mukhanov recently proposed a modified version of general relativity that implements the idea of a limiting curvature. In the spatially flat, homogeneous, and isotropic sector, their theory turns out to agree with the effective dynamics of the simplest version of loop quantum gravity if one identifies their limiting curvature with a multiple of the Planck curvature. 
At the same time, it extends to full general relativity without any symmetry assumptions and thus provides an ideal toy model for full loop quantum gravity in the form of a generally covariant effective action known to all orders. In this paper, we study the canonical structure of this theory and point out some interesting lessons for loop quantum gravity. We also highlight in detail how the two theories are connected in the spatially flat, homogeneous, and isotropic sector. 
\end{abstract}

}

\section{Introduction}

Quantum gravity is expected to challenge the traditional notion of Pseudo-Riemannian spacetime underlying general relativity and replace it with a more fundamental concept, in particular introducing a certain notion of ``fuzziness'' owed to quantum theory. Given any fundamental theory of quantum gravity, it is of great interest to extract an effective large scale theory from which observational consequences can be derived. Within loop quantum cosmology, this can be done using effective equations which have proven to accurately mimic the quantum evolution \cite{RovelliWhyAreThe}. For full loop quantum gravity, no effective equations are known, and the task to derive them appears to be technically extremely complicated. In this context, it would be very desirable to have four-dimensional toy models without any symmetry assumptions that still incorporate simple quantum corrections similar to those in loop quantum cosmology.  

In this paper, we would like to advocate a model recently proposed by Chamseddine and Mukhanov \cite{ChamseddineResolvingCosmologicalSingularities} that modifies general relativity with a limiting curvature as such a toy model for an effective theory of quantum gravity. In \cite{ChamseddineResolvingCosmologicalSingularities}, quantum gravity was not an objective (but a motivation via non-commutative geometry \cite{ChamseddineQuantaOfGeometryNoncommutative}) and the limiting curvature was chosen to be significantly below the Planck curvature in order to avoid the issue of particle production (which we are going to ignore in this paper). Here, we would like to take the viewpoint that the limiting curvature should be set to the Planck scale. We will in particular highlight that this makes sense by a quantum mechanical argument in section \ref{eq:FChi} and a direct comparison to spatially flat, homogeneous, and isotropic loop quantum cosmology in section \ref{sec:HomIso}. 
Since most of the recent work in understanding loop quantum gravity at an effective level has been done in the Hamiltonian framework, it is important to understand the canonical structure of the model proposed in \cite{ChamseddineResolvingCosmologicalSingularities}. This will be addressed in section \ref{sec:CanAn} and constitute the main part of the paper. Previous work on this matter includes \cite{MalaebHamiltonianFormulationOf, KlusonCanonicalAnalysisOf, FirouzjahiInstabilitiesInMimetic}.

\section{Canonical structure of GR with a limiting curvature} \label{sec:CanAn}

\subsection{Constraint analysis}

We start with the action \cite{ChamseddineResolvingCosmologicalSingularities}
\be
	S = \int d^4x \, \sqrt{-g} \left( \frac{1}{2} R + \frac{1}{2} \lambda \left(1+ g^{\mu \nu} \partial_\mu \phi \partial_\nu \phi \right) + f (\Box \phi)\right) \label{eq:MCAction}
\ee
where we have chosen the $(-, +, +, +)$ signature convention and $8 \pi G = 1$. $\lambda$ is a Lagrange multiplier field which enforces the constraint $g^{\mu \nu} \partial_\mu \phi \partial_\nu \phi = -1$. In the homogeneous and isotropic sector, it implies that $\Box \phi \propto \frac{\dot a}{a}$. $f$ is an a priori arbitrary function that can be fixed later to give simple equations of motion. It is constrained by the requirement that general relativity has to be obtained at low curvatures.

A canonical analysis of \eqref{eq:MCAction} has already been given in \cite{KlusonCanonicalAnalysisOf}. The form of the Dirac brackets was however only sketched and no explicit expressions were given. For our analysis later in the paper, we need those precise expressions and thus need to compute them. In order to remain self-contained, we will reiterate the constraint analysis from the beginning. We will deviate slightly from \cite{KlusonCanonicalAnalysisOf} in that we do not solve the equations of motion for $\chi$ (see below) before the analysis in order not to obtain only implicitly known functions. The Hamiltonian analysis of a related model, mimetic gravity, has already been performed in \cite{MalaebHamiltonianFormulationOf}. There are however strong differences in the analysis stemming from the substitution of $f(\Box \phi) \rightarrow f(\phi)$, which considerably simplifies the analysis and changes the resulting physics. See also \cite{FirouzjahiInstabilitiesInMimetic} for additional literature. 

Throughout this analysis, we will neglect any boundary terms. As usual, we restrict our spacetime to be globally hyperbolic so that we can perform a $3+1$-split. 
For the canonical analysis, the arbitrary function $f$ is problematic, since it depends on time derivatives of the canonical variables, in particular the metric. Therefore, we will use the equivalent action 
\begin{align}
	S &= \int d^4x \, \sqrt{-g} \left( \frac{1}{2} R + \frac{1}{2} \lambda \left(1+ g^{\mu \nu} \partial_\mu \phi \partial_\nu \phi \right) + f (\chi) + \beta (\chi - \Box \phi)\right) \nonumber \\
	&	= \int d^4x \, \sqrt{-g} \left( \frac{1}{2} R + \frac{1}{2} \lambda \left(1+ g^{\mu \nu} \partial_\mu \phi \partial_\nu \phi \right) + f (\chi) + \beta \chi + g^{\mu \nu} \partial_\mu \beta \partial_\nu \phi\right)
\end{align}
where we enforced $\chi = \Box \phi$ via the Lagrange multiplier field $\beta$ and then partially integrated. The dependence of this action on time derivatives of the metric is now the same as for the ADM action \cite{ArnowittTheDynamicsOf} and the canonical analysis can proceed as in this case. We obtain the canonical momenta
\begin{align}
	P^{ab} = &\frac{\partial \mathcal L}{\partial \dot q_{ab}} =  \frac{1}{2} \sqrt{q} \left(K^{ab} - K q^{ab} \right)\\
	P_\beta =& \frac{\partial \mathcal L}{\partial \dot \beta} =  \frac{\sqrt{q}}{N} \left( - \partial_t \phi +N^a \partial_a \phi \right) \\
	P_\phi =& \frac{\partial \mathcal L}{\partial \dot \phi} =   \frac{\lambda \sqrt{q}}{N} \left( - \partial_t \phi +N^a \partial_a \phi \right)+  \frac{\sqrt{q}}{N} \left( - \partial_t \beta +N^a \partial_a \beta \right)    
\end{align}
\begin{align}
	P_\lambda =& \frac{\partial \mathcal L}{\partial \dot \lambda} \approx 0\\
	P_\chi =& \frac{\partial \mathcal L}{\partial \dot \chi} \approx 0\\
	P_N =& \frac{\partial \mathcal L}{\partial \dot N} \approx 0\\
	P_{N^a} =& \frac{\partial \mathcal L}{\partial \dot N^a} \approx 0
\end{align}
where $q_{ab}$ denotes the spatial metric, $q$ its determinant, $K_{ab}$ the extrinsic curvature, $K = K_{ab} q^{ab}$ its trace, $N$ the lapse function, and $N^a$ the shift vector. 
$a, b, \ldots = 1,2,3$ are spatial tensor indices. $\approx$ is Dirac's notation \cite{DiracLecturesOnQuantum} for an equation valid on the constraint surface. Here, the last four equations are constraints and follow since the corresponding velocities cannot be expressed through their momenta.

After the Legendre transform, the Hamiltonian reads
\be
	H = \int d^3 x \, \left( N \mathcal H + N^a \mathcal H_a + \alpha_N P_N + \alpha_{N^a} P_{N^a} + \alpha_\lambda P_\lambda + \alpha_\chi P_\chi \right)
\ee
where $\alpha_N$, $\alpha_{N^a}$, $\alpha_\chi$, and $\alpha_\lambda$ are Lagrange multipliers and 
\begin{align}
	\mathcal H &= \frac{2}{\sqrt{q}} P^{ab} P^{cd} \left(q_{ac} q_{bd} - \frac{1}{2} q_{ab} q_{cd} \right) - \frac{1}{2} \sqrt{q} R^{(3)}  - \frac{1}{2} \lambda \sqrt{q} \left( 1+ q^{ab} \partial_a \phi \partial_b \phi - \frac{P_\beta^2}{q}\right) \nonumber \\
		&~~~~ - \sqrt{q} \left( f(\chi) + \beta \chi \right) - \sqrt{q} q^{ab} \partial_a \beta \partial_b \phi - \frac{1}{\sqrt{q}} P_\phi P_\beta \\
	\mathcal H_a &= - 2 \nabla_b P^{b} {}_a + P_\beta \partial_a \beta + P_\phi \partial_a \phi + P_\chi \partial_a \phi + P_\lambda \partial_a \lambda \text{.} \label{eq:Ha}
\end{align}
The last two terms in \eqref{eq:Ha} have been added by redefining $\alpha_\chi$ and $\alpha_\lambda$ in order to make $\mathcal H_a$ first class. This is always possible due to $P_\lambda \approx 0$ and $P_\chi \approx 0$. The non-vanishing Poisson brackets read
\begin{align}
	\left\{q_{ab}(x), P^{cd}(y) \right\} &= \delta^{(3)}(x,y) \delta_{(a}^c \delta_{b)}^d\\
	\left\{\beta(x), P_\beta (y) \right\} &= \delta^{(3)}(x,y) \\
	\left\{\phi(x), P_\phi (y) \right\} &= \delta^{(3)}(x,y) \\
	\left\{\lambda(x), P_\lambda (y) \right\} &= \delta^{(3)}(x,y) \\
	\left\{\chi(x), P_\chi (y) \right\} &= \delta^{(3)}(x,y) \\
	\left\{N(x), P_N (y) \right\} &= \delta^{(3)}(x,y) \\
	\left\{N^a(x), P_{N^b} (y) \right\} &= \delta^{(3)}(x,y) \delta^a_b \text{,}
\end{align}
where the symmetrisation is defined as $f_{(a} g_{b)} = (f_a g_b + f_b g_a)/2$.

As a next step, we need to ensure stability of the constraints obtained in the singular Legendre transform. From $\dot P_N = \{ P_N, H\} = - \mathcal H  \stackrel{!}{\approx} 0$ and $\dot P_{N^a} = \{ P_{N^a}, H\} = - \mathcal H_a  \stackrel{!}{\approx} 0$ we obtain the Hamiltonian constraint $\mathcal H$ and the spatial diffeomorphism constraint $\mathcal H_a$. Next, we compute
\be
	\dot P_\lambda = \{ P_\lambda, H\} = \frac{N \sqrt{q}}{2}  \left( 1+ q^{ab} \partial_a \phi \partial_b \phi - \frac{P_\beta^2}{q}\right) =:  \frac{N \sqrt{q}}{2} C_\lambda  \stackrel{!}{\approx} 0
\ee
and add $C_\lambda = 1+ q^{ab} \partial_a \phi \partial_b \phi - P_\beta^2 / {q} \approx 0$ to the list of constraints. We also obtain
\be
	\dot P_\chi = \{ P_\chi, H\} = {N \sqrt{q}} \left( \beta + f'(\chi) \right) =:  {N \sqrt{q}} \, C_\chi \stackrel{!}{\approx} 0
\ee
and add $C_\chi = \beta + f'(\chi)$ to the list of constraints. 

The same analysis now needs to be reiterated to check consistency of the new constraints. We start with 
\begin{align}
	\dot C_\lambda = \{ C_\lambda, H\} \approx - N \frac{2}{\sqrt{q}} \Bigg[ & 2 P^{ab} \partial_a \phi \partial_b \phi + P \left( \frac{P_\beta^2}{q}  - q^{ab} \partial_a \phi \partial_b \phi \right) + P_\beta \chi - \frac{P_\beta}{\sqrt{q}} \partial_a \left( \sqrt{q} q^{ab} \partial_b \phi \right) \nonumber \\
	&+ \sqrt{q} q^{ab} \left( \partial_a \frac{P_\beta}{\sqrt{q}}\right) \partial_b \phi \Bigg]  \stackrel{!}{\approx} 0
\end{align}
where $P = P^{ab} q_{ab}$. Using $C_\lambda \approx 0 ~~ \Leftrightarrow ~~ P_\beta = \sqrt{q} \w$, we obtain the new constraint 
\begin{align}
	D_\lambda = &\chi + \frac{2 P^{ab} \partial_a \phi \partial_b \phi }{\sqrt{q} \w} + \frac{P}{\sqrt{q} \w}- \Delta \phi  + {q^{ab} \partial_a \phi}\partial_b \log \w \approx 0
\end{align}
where $\Delta \phi = \frac{1}{\sqrt{q}} \partial_a \left( \sqrt{q} q^{ab} \partial_b \phi \right)$. Stability of $C_\chi \approx 0$ under time evolution can be ensured by fixing the Lagrange multiplier $\alpha_\chi$. The details of this computation are irrelevant for the current paper and we will not spell them out here. Since $\mathcal H_a$ generates spatial diffeomorphisms, it is trivially stable under time evolution generated by an integrated density, such as $H$. $\mathcal H$ also turns out to be stable due to $\{ \mathcal H[M], \mathcal H[N]\} = \mathcal H_a [q^{ab} \left( M \partial_b N - N \partial_b M \right)]$, which requires only little more computation than in the GR case. 
Finally, we have to check stability of $D_\lambda$. $\dot D_\lambda = \{ D_\lambda, H\}  \stackrel{!}{\approx} 0$ yields an equation for $\lambda$, which again has to be added to the list of constraints. The details are again not important for what follows and we write the constraint as $E_\lambda = \lambda + \ldots \approx 0$. At last, the stability of $E_\lambda$ can be ensured by fixing $\alpha_\lambda$. 

The Dirac algorithm ends here. To simplify the bookkeeping a little, we can solve $P_N \approx 0$ and $P_{N^a} \approx 0$ by promoting $N$ and $N^a$ to Lagrange multipliers. Also, we can solve $E_\lambda \approx 0$ and $P_\lambda \approx 0$ by setting $P_\lambda = 0$ and substituting $\lambda$ in $\mathcal H$ accordingly. 
For the remaining constraints, we are interested in splitting them into first and second class subsets. It turns out that $P_\chi, D_\lambda, C_\lambda$, and $C_\chi $ form a second class subset, since their Dirac Matrix is invertible (see the next subsection). $\mathcal H$ and $\mathcal H_a$ are Poisson-commuting among themselves on the constraint surface and can be made to Poisson-commute also with the second class constraints by adding suitable correction terms to $\mathcal H$. Since we are interested in using the Dirac bracket later on, we will not compute these correction terms, as they will drop out in the Dirac bracket anyway. 

To summarise, we are left with the first class constraints $\mathcal H$ and $\mathcal H_a$ (up to said correction terms), and the second class constraints $P_\chi, D_\lambda, C_\lambda$, and $C_\chi $. The remaining canonical pairs are $\{q_{ab}, P^{cd}\}$, $\{ \beta, P_\beta\}$, $\{ \phi, P_\phi\}$, and $\{ \chi, P_\chi\}$. The theory thus has one more degree of freedom than general relativity, which should be identified with the canonical pair $\{ \phi, P_\phi\}$ providing a Dark matter candidate \cite{ChamseddineMimeticDarkMatter}.

\subsection{Dirac bracket}

The main virtue of the Dirac bracket is that it implements the second class constraints of a Hamiltonian system, i.e. they can be imposed either before or after the Dirac brackets have been evaluated. In order to construct them, we assemble the second class constraints in a vector $S_i = (P_\chi, D_\lambda, C_\lambda, C_\chi )_i$ and compute the Dirac matrix
\be
	M_{ij}(x,y) = \left\{ S_i(x), S_j(y) \right\} = \begin{pmatrix}
0 & -1 & 0 & -f''(\chi) \\
1 & \gamma & -\frac{2}{\sqrt{q}}\left( 2 w-  \frac{1}{2 w} \right) & 0 \\
0 & \frac{2}{\sqrt{q}}\left( 2 w-  \frac{1}{2 w} \right)  & 0 & \frac{2w}{\sqrt{q}} \\
f''(\chi) & 0 & -\frac{2w}{\sqrt{q}} & 0
\end{pmatrix}_{ij} \delta^{(3)}(x,y)
\ee
where we abbreviated $w := \w$ and $\gamma = \frac{w^2-3/2}{\sqrt{q}w} \left( \partial^a  \phi \right)(y) \stackrel{(x)}{\nabla_a} - \frac{w^2-3/2}{\sqrt{q}w} \left( \partial^a  \phi \right)(x) \stackrel{(y)}{\nabla_a} $, so that $\gamma$ encodes contributions to the Dirac matrix that are proportional to derivatives of the $\delta$-distribution (note that, e.g., $\stackrel{(x)}{\nabla_a}$ acts only on the $x$-argument of  $\delta^{(3)}(x,y)$.). We see that the resulting Dirac bracket will be rather cumbersome, although straight forwardly computable. A huge simplification occurs if we restrict us to spatial slices where $\partial_a \phi = 0$, i.e. $\phi = \phi(t)$. Such a restriction can be seen as a gauge choice for the Hamiltonian constraint which forces us to set $N = N(t)$ for consistency. Since the main points we want to make in this paper are not affected by such a restriction, we are going to assume $\partial_a \phi = 0$ in the following {\it after} evaluating Poisson or Dirac brackets. We will remind ourselves about this fact by writing equalities using $\stackrel{\partial_a \phi=0}{=}$. 

The Dirac matrix can be inverted to 
\be
	\left( M^{-1} \right)^{ij} (x,y) \stackrel{\partial_a \phi=0}{=} \frac{\sqrt{q} \delta^{(3)}(x,y)}{3 f''(\chi) -2 } \begin{pmatrix}
0 & - \frac{2}{\sqrt{q}}  & 0 & \frac{3}{\sqrt{q}}  \\
\frac{2}{\sqrt{q}} & 0 & f''(\chi) & 0 \\
0 & -f''(\chi) & 0 & 1 \\
-\frac{3}{\sqrt{q}}  & 0 & -1 & 0
\end{pmatrix}^{ij} \text{.}
\ee
The Dirac brackets between two phase space functions $A$ and $B$ are defined as
\be
	\{A, B\}_* = \{A, B\} - \sum_{i,j} \int d^3x d^3y \, \{A, S_i(x)\} \left( M^{-1} \right)^{ij}(x,y) \{ S_j(y), B\} \text{.}
\ee
As an example, we compute 
\be
	\left\{q_{ab}(x), P^{cd}(y) \right\}_* \stackrel{\partial_a \phi=0}{=} \delta^{(3)}(x,y) \left( \delta_{(a}^c \delta_{b)}^d - q_{ab} q^{cd} \frac{f''(\chi)}{3 f''(\chi) -2} (x ) \right)
\ee
The full set of non-vanishing Dirac brackets is given by appendix \ref{app:DB}.

An interesting choice (further discussed in section \ref{eq:FChi}) for $f(\chi)$ proposed in \cite{ChamseddineResolvingCosmologicalSingularities} that ultimately leads to simple dynamics is (adapted to our sign conventions) given by
\be
	f(\chi) = - \chi_m^2 \, g\left( \sqrt{\frac{2}{3}} \frac{\chi}{\chi_m} \right), ~~~ g(y) = -1 -\frac{y^2}{2} + y \arcsin y + \sqrt{1-y^2} \label{eq:f}
\ee
where $\chi_m$ is a constant determining the limiting curvature. This yields
\begin{align}
	\left\{q_{ab}(x), P^{cd}(y) \right\}_* & \stackrel{\partial_a \phi=0}{=} \delta^{(3)}(x,y) \left( \delta_{(a}^c \delta_{b)}^d - \frac{1}{3}q_{ab} q^{cd} \left(1-\sqrt{1-\frac{2}{3} \frac{\chi^2}{\chi_m^2}} \right)\right) \nonumber \\
		& ~ ~=~~ \delta^{(3)}(x,y) \left( \frac{1}{3}q_{ab} q^{cd} \sqrt{1-\frac{2}{3} \frac{\chi^2}{\chi_m^2}} + \left( \delta_{(a}^c \delta_{b)}^d  - \frac{1}{3}q_{ab} q^{cd} \right)  \right) \text{,} \label{eq:DBqP}
\end{align}
showing that brackets involving the trace part of $P^{ab}$ are deformed as opposed to standard general relativity.

\subsection{Gauge transformations and algebra}

The algebra of the remaining constraints $\mathcal H$ and $\mathcal H_a$ now has to be evaluated using the Dirac bracket. 
We will continue to do this under the simplification $\partial_a \phi = 0$ {\it after} evaluating Dirac brackets.
The result (it does not change) can already be anticipated by an appeal to general covariance, see for example \cite{HojmanGeometrodynamicsRegained} for an overview: the algebra of constraints has to reflect the fact that we are dealing with a generally covariant theory in four spacetime dimensions and thus only the standard hypersurface deformation algebra can follow: 
\begin{align}
	\left\{ \mathcal H_a[M^a], \mathcal H_b[N^b]\right\}_* &\stackrel{\partial_a \phi=0}{=}  \mathcal H_a[[M,N]^a]\\
	\left\{ \mathcal H_a[M^a], \mathcal H[N] \right\}_* &\stackrel{\partial_a \phi=0}{=}  \mathcal H[\mathcal L_M N]\\
	\left\{ \mathcal H[M], \mathcal H[N] \right\}_* &\stackrel{\partial_a \phi=0}{=}  \mathcal H_a[q^{ab} \left( M\partial_b  N - N \partial_b M \right)]
\end{align} 
In the computation, several possible deformation terms in the algebra cancel after adding up all terms, reflecting the underlying four-diffeomorphism symmetry of the action. For example, one obtains terms of the form
\be
	\pm \frac{1}{3} \left( \mathcal L_M P^{ab} \right) \left( \mathcal L_n q_{cd}  \right) q_{ab} q^{cd}  \left(1-\sqrt{1-\frac{2}{3} \frac{\chi^2}{\chi_m^2}}\right)
\ee
 twice with opposite signs in the bracket of two diffeomorphism constraints and
\be
	\pm \frac 23 \nabla^{(3)}_c \left( q^{cd} P\right) \left( M\partial_b  N - N \partial_b M \right)  \left(1-\sqrt{1-\frac{2}{3} \frac{\chi^2}{\chi_m^2}}\right)
\ee
in the bracket of two Hamiltonian constraints. 
The general case $\partial_a \phi \neq 0$ is extremely laborious and does not promise to yield additional insights. Due to the above argument, we expect the algebra to be undeformed.

On the other hand, it is interesting to investigate the action of the spatial diffeomorphism constraint on the phase space variables, here again for the case $\partial_a \phi = 0$, but for general $f(\chi)$. We obtain
\begin{align}
	\{q_{ab}, \mathcal H_a [N^a]\}_*  &\stackrel{\partial_a \phi=0}{=} \mathcal L_N q_{ab} - \left( \mathcal L_N q_{cd} \right)q^{cd} q_{ab} \frac{f''(\chi)}{3f''(\chi)-2}  + \left( \mathcal L_N P_\beta \right) \frac{q_{ab}}{\sqrt{q}}  \frac{2f''(\chi)}{3f''(\chi)-2} \nonumber\\
		&  \stackrel{\partial_a \phi=0}{=}  \mathcal L_N q_{ab} \label{eq:qHa}\\
	\{P^{ab}, \mathcal H_a [N^a]\}_*  &\stackrel{\partial_a \phi=0}{=} \mathcal L_N P^{ab} - \mathcal L_N \left( \frac{P}{\sqrt{q}} \right) \sqrt{q} q^{ab} \frac{f''(\chi)}{3f''(\chi)-2} + \left( \mathcal L_N \beta \right) \sqrt{q} q^{ab} \frac{1}{3f''(\chi)-2} \nonumber \\
		&  \stackrel{\partial_a \phi=0}{=}  \mathcal L_N P^{ab}  \label{eq:DiffonPDB} \text{.}
\end{align}
where we used $C_\lambda = 0$ in \eqref{eq:qHa} as well as $D_\lambda = 0$ and $C_\chi = 0$ in \eqref{eq:DiffonPDB}. 
We see that both $q_{ab}$ and $P^{ab}$ are Lie dragged due to cancellations of all other involved terms. In addition, one can show that also all other phase space variables are Lie dragged due to similar cancellations. This provides an independent check for the algebra relations involving the generator of spatial diffeomorphisms.

\section{Relation to loop quantum gravity} \label{sec:LQCRelation}

\subsection{Spatially flat, homogeneous, and isotropic sector} \label{sec:HomIso}

The spatially flat, homogeneous and isotropic sector of the theory defined by the action \eqref{eq:MCAction} has been studied in \cite{ChamseddineResolvingCosmologicalSingularities}. It leads to the equation
\be
	3 \left( \frac{\dot a}{a} \right)^2 = \frac{\epsilon_m}{a^{3(1+\omega)}} \left(1 - \frac{1}{a^{3(1+\omega)}} \right) \label{eq:EOMHOMMC}
\ee
where $a$ is the (suitably normalised) scale factor, $\omega$ determines the equation of state $p = \omega \epsilon$,  $\epsilon_m = 2 \chi^2_m$, and and we neglected the contribution from the mimetic dark matter. Its solution is given by 
\be
	a(t) = \left(1+ \frac{3}{4}(1+\omega)^2 \epsilon_m t^2 \right)^{\frac{1}{3(1+\omega)}} \text{.}
\ee
Both the equation and its solution are known also from loop quantum gravity, more precisely loop quantum cosmology\footnote{See \cite{BodendorferStateRefinementsAnd, OritiEmergentFriedmannDynamics} and references therein for recent progress on deriving loop quantum cosmology from loop quantum gravity.}, see for example equations (2)-(6) in \cite{Wilson-EwingTheMatterBounce} (after appropriately rescaling the scale factor and setting $8 \pi G=1$, we have $\rho_c = \epsilon_m$, where $\rho_c$ is the critical energy density at which the universe bounces.).

Alternatively, one can take a look at the effective action of spatially flat, homogeneous, and isotropic loop quantum cosmology, which has been computed in \cite{DateEffectiveActionsFrom} by a Legendre-transform of the effective Hamiltonian theory. The choice \eqref{eq:f} for the function $f(\chi)$ can be directly read off from equation (7) of \cite{DateEffectiveActionsFrom}, see also \cite{HellingHigherCurvatureCounter} equation (16). The mimetic dark matter contribution present in \eqref{eq:MCAction} is missing in loop quantum cosmology, but it can in principle be added as an additional matter field. 

In yet another way, equivalence can be established in the Hamiltonian formulation. We will do this here to test our results from the previous section. Reduced to the homogeneous and isotropic sector, \eqref{eq:DBqP} becomes
\be
	\left\{ v, \chi \right\}_* = - \frac{3}{2} \sqrt{1-\frac{2}{3} \frac{\chi^2}{\chi_m^2}} \label{eq:PBMCRED}
\ee
where $v = a^3$ (with unit fiducial volume) and $\chi = 3 \frac{\dot a}{a}$. 
The gravitational part of the Hamiltonian constraint reduces to 
\be
	\mathcal H_\text{grav} = - \chi_m^2 v \left(1- \sqrt{1-\frac{2}{3} \frac{\chi^2}{\chi_m^2}}\right) \text{.} \label{eq:HMCRED}
\ee
\eqref{eq:EOMHOMMC} can now be obtained from $\mathcal H_\text{grav} + \mathcal H_\text{matter}=0$ by using \eqref{eq:PBMCRED} to identify $\chi = 3 \frac{\dot a}{a}$. To simplify the Dirac bracket \eqref{eq:PBMCRED}, we can use the variable transformation
\be
	b = - \sqrt{\frac{3}{2}} \, \chi_m \, \arcsin \left( \sqrt{\frac{2}{3}} \frac{\chi}{\chi_m} \right) \text{.}
\ee
The Dirac bracket now simplifies to 
\be
		\left\{ v, b \right\}_* =  \frac{3}{2} 
\ee 
and the gravitational part of the Hamiltonian becomes
\be
	\mathcal H_\text{grav} = - \frac{1}{3} \left(\sqrt{6} \chi_m \right)^2 v \sin^2 \left( \frac{b}{\sqrt{6} \chi_m} \right) =: - \frac{1}{3} \frac{1}{\lambda^2} v \sin^2 \left( \lambda b \right) \text{.} \label{eq:HGRAVLQC}
\ee
\eqref{eq:HGRAVLQC} is however nothing else than the gravitational part of the effective loop quantum cosmology Hamiltonian constraint, see \cite{SinghLoopQuantumCosmologyABrief} for a review, with the polymerisation scale $\lambda$ identified as $\frac{1}{\sqrt{6} \chi_m}$.

\subsection{Choice for $f(\chi)$} \label{eq:FChi}

The form of $f(\chi)$ has been guessed in \cite{ChamseddineResolvingCosmologicalSingularities} with the aim to simplify the equations of motion, which requires considerable effort. $f(\chi)$ can however be derived by a simple quantum mechanical argument as follows. In \cite{ChamseddineResolvingCosmologicalSingularities}, the motivation given for the form of the action was non-commutative geometry, in particular the quantisation of three-volume \cite{ChamseddineQuantaOfGeometryNoncommutative} in equidistant steps (integer steps in the following). If the three-volume is quantised as such, its canonical conjugate, the mean curvature $b$, cannot exist as an operator, as it would act as a derivative on wave functions $\psi(v)$ which have support only on discrete values of $v$. Rather, the exponential $e^{i n b}$ for $n \in \mathbb Z$ can exist as an operator. This necessitates to approximate $b$ by some function of $e^{i n b}$, the simplest of which is $\sin b$. This directly leads to the effective Hamiltonian of loop quantum cosmology, see section \ref{sec:HomIso}, from which $f(\chi)$ can be obtained by a Legendre transform as in \cite{DateEffectiveActionsFrom}, equation (7). The limiting curvature is then naturally identified with an order 1 multiple of the Planck curvature.  

We note that this is the usual logic of replacing connections by holonomies used in loop quantum gravity. In the current context, using wave functions depending on U$(1)$ (point)-holonomies $e^{i K}$ of the mean curvature $K = K_{ab} q^{ab}$ would lead to the equidistant spacing of volume eigenvalues \cite{BVI}. The unifying concept behind the non-commutative geometry motivation of \cite{ChamseddineResolvingCosmologicalSingularities} and loop quantum cosmology that suggests the form of $f(\chi)$ is thus the equidistant spectrum of the volume operator.

\subsection{Conceptual remarks} 

As soon as one leaves the homogeneous and isotropic sector, the effective dynamics of loop quantum cosmology starts to deviate from those of \eqref{eq:MCAction}. This can be seen simply in the different way the two theories receive higher curvature corrections. Whereas in homogeneous but non-isotropic loop quantum cosmology \cite{AshtekarLoopQuantumCosmologyBianchi} one would polymerize (replace $x \mapsto \frac{1}{\lambda} \sin (\lambda x) $) the three independent components of the extrinsic curvature seperately, one only adds corrections proportional to the trace of the extrinsic curvature in \eqref{eq:MCAction}. In the general non-symmetric case, the corrections are more complicated and also involve spatial derivatives of the scalar fields, however those are not present in loop quantum gravity to begin with. We have also explicitly checked by inserting the homogeneous but non-isotropic solution for \eqref{eq:MCAction} into the effective Hamiltonian of \cite{AshtekarLoopQuantumCosmologyBianchi} that the two theories deviate in the details. However, the qualitative features of the numerical analysis of \cite{AshtekarLoopQuantumCosmologyBianchi} agree with the exact solution given in \cite{ChamseddineResolvingCosmologicalSingularities}: Kasner exponents smoothly change while traversing the high curvature regime with qualitatively similar transition rules. 

The question of closure or deformation of the constraint algebra discussed in the previous section has recently been an important subject of debate in canonical loop quantum gravity. Since one is working in the canonical framework, one introduces quantum corrections in a $3+1$ split of spacetime and has to later check if the quantum analogue of the hypersurface deformation algebra is broken or deformed. Most noticeably, it was found in specific symmetry reduced models \cite{CailleteauConsistencyOfHolonomy} that a deformation of the type
\be
	\left\{ \mathcal H[M], \mathcal H[N] \right\} = \beta \,  \mathcal H_a[q^{ab} \left( M\partial_b  N - N \partial_b M \right)] \label{eq:HH}
\ee
occurs, where $\beta$ (independent of the $\beta$ in section \ref{sec:CanAn}) is a function that transitions from $1$ in low curvatures to $-1$ at the critical energy density. This deformation has been interpreted as an effective signature change from a Lorentzian to a Euclidean regime, see \cite{BojowaldSomeImplicationsOf, BojowaldHypersurfaceDeformationAlgebroids} for an overview. It was later found that this effect is absent, i.e. $\beta = 1$, if one works with self-dual variables (and the associated quantum corrections) \cite{AcourSphericallySymmetricSector, AchourANewLook}, which are distinguished from the point of view that they can be seen as the pullbacks of spacetime-covariant quantities. One would thus conclude that the answer to the question of algebra deformation is a result of the type of quantum corrections one introduces, i.e. whether they are spacetime-covariant or not. Using self-dual variables, one naturally has a better chance of maintaining spacetime covariance. 

This connects to the present paper as follows.
In order to check \eqref{eq:HH}, one needs to go beyond the homogeneous and isotropic sector and at least introduce cosmological perturbations. Therefore, the quantum corrections in the spatially flat, homogeneous and isotropic sector considered in this section are not enough to determine whether the algebra is deformed. However, \eqref{eq:MCAction} is well defined without any symmetry assumptions and in particular {\it fully spacetime-covariant}. Therefore, \eqref{eq:MCAction} already prescribes a way to implement spacetime-covariant quantum corrections for cosmological perturbations. It would be interesting to compare these (see \cite{FirouzjahiInstabilitiesInMimetic}) to those in \cite{AchourANewLook} and to draw general lessons for how to achieve spacetime covariance when polymerising in the Hamiltonian framework.

While the spatial diffeomorphism constraint has been deformed, its action on the canonical variables via a Lie derivative remained unchanged due to the deformed phase space. 
For loop quantum gravity, this is of strong interest since one usually implements the spatial diffeomorphism constraint using the classical flow \cite{AshtekarQuantizationOfDiffeomorphism} instead of polymerising it first on the same footing as the Hamiltonian constraint. \eqref{eq:DiffonPDB} suggests that this strategy is self-consistent with an alternative route via polymerisation, supporting the findings of \cite{LaddhaTheDiffeomorphismConstraint}. The deformation of phase space and the form of the spatial diffeomorphism constraint should also be relevant for the question of the closure of the quantum constraint algebra. 

A deformed spatial diffeomorphism constraint has also been observed in \cite{AchourANewLook} in the context of cosmological perturbations, where the spatial diffeomorphism constraint was also polymerised, while the constraint algebra remained undeformed. However, in \cite{AchourANewLook}, the preliminary conclusion was that the quantum flow might deviate from the classical one. In our computation, we saw to the contrary that the interplay of a polymerised constraint with a deformed Dirac bracket can again result in an undeformed flow.

\section{Conclusion}

In this paper, we have discussed the canonical structure of the action \eqref{eq:MCAction}, which modifies general relativity with a limiting curvature. In order to perform the analysis, we have rewritten the action slightly to bring the gravitational sector in the form of general relativity. This lead to second class constraints coding the higher curvature corrections, which we have implemented using the method of Dirac brackets. The constraint algebra has been shown to remain in standard form in the simplified case $\partial_a \phi=0$ and was argued to remain so also in the general case. On top of the previous analysis \cite{KlusonCanonicalAnalysisOf}, we added explicit expressions for the Dirac bracket and computed the action of the spatial diffeomorphism constraint on the phase space on a surface of constant scalar field $\phi$. We found that the flow generated by the spatial diffeomorphism constraint is the standard one due to cancellations between deformations in the constraint and the Dirac bracket. 
While we restricted to four spacetime dimensions in this paper, there do not seem to be any difficulties in extending the current computations to higher dimensions. 

We have discussed in detail how \eqref{eq:MCAction} is equivalent to the effective action of loop quantum cosmology in the spatially flat, homogeneous and isotropic sector up to the additional ``mimetic'' matter contribution. The main virtue of the action \eqref{eq:MCAction} for loop quantum gravity is thus that it provides us with a toy model effective action that shares important features of loop quantum cosmology and is defined without any symmetry assumptions. 
At the same time, the equations are somewhat simpler than in effective loop quantum gravity beyond the homogeneous and isotropic sector and allow for a broader class of analytic solutions \cite{ChamseddineResolvingCosmologicalSingularities, ChamseddineNonsingularBlackHole}. 
Having access to the effective action is in particular interesting in the context of using loop quantum gravity for AdS/CFT, see for example \cite{BodendorferHolographicSignaturesOf} and references therein, where the gravitational on-shell action determines the generating functional of connected correlation functions in the dual field theory. We also gave an argument for the choice of the function $f(\chi)$ based on quantum mechanics with an equidistant volume spectrum. Since \cite{ChamseddineResolvingCosmologicalSingularities} was motivated by such an equidistant spectrum originating from non-commutative geometry \cite{ChamseddineQuantaOfGeometryNoncommutative}, this provides a sound basis for the choice made in \cite{ChamseddineResolvingCosmologicalSingularities}.

\section*{Acknowledgements}
NB was supported by a Feodor Lynen Return Fellowship of the Alexander von Humboldt-Foundation. Discussions with Viatcheslav Mukhanov and Edward Wilson-Ewing are gratefully acknowledged. 
While writing up our findings, we became aware of the simultaneous preparation of \cite{NouiEffectiveLoopQuantum}, which also points out the relation between \cite{ChamseddineResolvingCosmologicalSingularities} and loop quantum gravity. We also thank Karim Noui for pointing out reference \cite{KlusonCanonicalAnalysisOf} to us.

\begin{appendix}

\section{Non-vanishing Dirac brackets} \label{app:DB}
In this appendix, we explicitly compute all non-vanishing Dirac brackets in the case $\partial_a\phi = 0$. 
\begin{align}
	\left\{q_{ab}(x), P^{cd}(y) \right\}_* &= \delta^{(3)}(x,y) \left( \delta_{(a}^c \delta_{b)}^d - q_{ab}  q^{cd} \frac{f''(\chi) }{3f''(\chi) -2}  \right)\\
	\left\{P^{ab}(x), P^{cd} (y)\right\}_* &= \delta^{(3)}(x,y)\left( q^{ab} P^{cd} -q^{cd} P^{ab} \right)\frac{ -f''(\chi)}{3 f''(\chi)-2} \\
	\left\{\beta (x), P_\beta(y) \right\}_* &= \delta^{(3)}(x,y)	\frac{3 f''(\chi)}{3 f''(\chi)-2}\\
	\left\{\phi (x), P_\phi(y) \right\}_* &= \delta^{(3)}(x,y)	\\
	\left\{\beta (x), P_\phi(y) \right\}_* &= \frac{2 f''(\chi)}{3 f''(\chi)-2}(y)   \stackrel{(x)}{\nabla_c}\stackrel{(x)}{\nabla^c} \delta^{(3)}(x,y) \\
	\left\{\chi(x), P_\phi(y) \right\}_* &= \frac{-2 }{3 f''(\chi)-2}(y)   \stackrel{(x)}{\nabla_c}\stackrel{(x)}{\nabla^c} \delta^{(3)}(x,y) \\
	\left\{\chi(x), P_\beta (y)\right\}_* &=  \delta^{(3)}(x,y) \frac{-3}{3 f''(\chi)-2}\\
	\left\{q_{ab}(x), \beta(y) \right\}_* &= \delta^{(3)}(x,y) q_{ab} \frac{-2 f''(\chi) /\sqrt{q}}{3 f''(\chi) -2} \\
	\left\{q_{ab}(x), \chi(y) \right\}_* &= \delta^{(3)}(x,y)  q_{ab}  \frac{2/\sqrt{q}}{3 f''(\chi) -2} \\
	\left\{\beta (x), P^{ab}(y)\right\}_* &=\delta^{(3)}(x,y)  \frac{-2 f''(\chi)/ \sqrt{q}}{3 f''(\chi)-2}  \left(   P^{ab}  -\frac{1}{2} P  q^{ab}  \right)   \\
	\left\{P^{ab}(x), P_\beta (y)\right\}_* &= \delta^{(3)}(x,y) q^{ab}  \frac{ \sqrt{q}}{3 f''(\chi) -2} \\
	\left\{\chi (x), P^{ab}(y)\right\}_* &= \delta^{(3)}(x,y)  \frac{2/ \sqrt{q}}{3 f''(\chi)-2}  \left(   P^{ab}  -\frac{1}{2} P  q^{ab}  \right) \\
	\left\{P_\phi (x), P^{ab}(y) \right\}_* &=- \sqrt{q} q^{ab} \frac{f''(\chi)}{3 f''(\chi)-2}(y) \stackrel{(x)}{\nabla_c}\stackrel{(x)}{\nabla^c} \delta^{(3)}(x,y)
\end{align}

\end{appendix}


\raggedright

\end{document}